\documentclass[aps,twocolumn,floatfix,superscriptaddress,prl]{revtex4-2}
\usepackage{mathrsfs}
\usepackage{color}
\usepackage{graphicx}           
\usepackage{dcolumn}            
\usepackage{bm}                 
\usepackage{hyperref}           
\usepackage[latin1]{inputenc}   
\usepackage{pstricks}           
\usepackage{amsmath,amssymb}
\usepackage{version}
\usepackage{epstopdf}
\usepackage{lipsum}
\DeclareGraphicsExtensions{.png,.eps,.ps}
\usepackage{ulem}
\usepackage{float}
\usepackage{booktabs}
\usepackage{subfigure}
\usepackage{graphicx}
\usepackage{dcolumn}
\usepackage{bm}
\usepackage{braket}             
\usepackage{footmisc}

\usepackage{float}

\definecolor{red}{rgb}{1,0,0}

\begin{document}

	\title{Dynamical phase evolution of Coulomb-focused electrons in strong-field ionization probed by a standing light wave}
	
	\author{Yuan Gu}\email{These authors contribute equally to this work.}\address{School of Physics, Zhejiang Key Laboratory of Micro-Nano Quantum Chips and Quantum Control, Zhejiang University, Hangzhou, 310058, China}
	\author{Hao Liang}\email{These authors contribute equally to this work.}\address{Max Planck Institute for the Physics of Complex Systems, Dresden, 01187, Germany}
	\author{Weiran Zheng}\address{School of Physics, Zhejiang Key Laboratory of Micro-Nano Quantum Chips and Quantum Control, Zhejiang University, Hangzhou, 310058, China}
	\author{Aofan Lin}\address{School of Physics, Zhejiang Key Laboratory of Micro-Nano Quantum Chips and Quantum Control, Zhejiang University, Hangzhou, 310058, China}
	\author{Jiaye Zhang}\address{School of Physics, Zhejiang Key Laboratory of Micro-Nano Quantum Chips and Quantum Control, Zhejiang University, Hangzhou, 310058, China}
	\author{Zichen Li}\address{School of Physics, Zhejiang Key Laboratory of Micro-Nano Quantum Chips and Quantum Control, Zhejiang University, Hangzhou, 310058, China}    
	\author{Juan Du}\address{School of Physics, Zhejiang Key Laboratory of Micro-Nano Quantum Chips and Quantum Control, Zhejiang University, Hangzhou, 310058, China} 
	\author{Lei Ying}\address{School of Physics, Zhejiang Key Laboratory of Micro-Nano Quantum Chips and Quantum Control, Zhejiang University, Hangzhou, 310058, China} 
	\author{Peilun He}\address{State Key Laboratory of Dark Matter Physics, Key Laboratory for Laser Plasmas (Ministry of Education) and School of Physics and Astronomy, Collaborative Innovation Center for IFSA (CICIFSA), Shanghai Jiao Tong University, Shanghai 200240, China}     
	\author{Jan-Michael Rost}\address{Max Planck Institute for the Physics of Complex Systems, Dresden, 01187, Germany}        
	\author{ Sina Jacob}\address{Institut f\"ur Kernphysik, Goethe-Universit\"at Frankfurt am Main, Frankfurt am Main 60438, Germany}         
	\author{Maksim Kunitski}\address{Institut f\"ur Kernphysik, Goethe-Universit\"at Frankfurt am Main, Frankfurt am Main 60438, Germany}  
	\author{Till Jahnke}\address{Max-Planck-Institut f\"ur Kernphysik, Heidelberg, 69117, Germany}
	\author{Sebastian Eckart}\address{Institut f\"ur Kernphysik, Goethe-Universit\"at Frankfurt am Main, Frankfurt am Main 60438, Germany}    
	\author{Kang Lin}\email{klin@zju.edu.cn}\address{School of Physics, Zhejiang Key Laboratory of Micro-Nano Quantum Chips and Quantum Control, Zhejiang University, Hangzhou, 310058, China}       
	\author{Reinhard D\"orner}\address{Institut f\"ur Kernphysik, Goethe-Universit\"at Frankfurt am Main, Frankfurt am Main 60438, Germany}

	\date{\today}

	\begin{abstract}
		{
			We investigate the dynamical phase evolution of Coulomb-focused electrons in strong-field ionization. We diffract the electrons with an ultrashort standing light wave to track their time-dependent phase. Our findings show that low-energy electrons exhibit a unique chromosome-shaped diffraction pattern, distinct from higher-energy electrons. Our numerical model quantitatively reproduces the experimental results, confirming this pattern maps the electron's time-dependent phase evolution as it escapes from a Coulomb potential. Our pulsed diffraction grating technique offers a new way to sense an electron's quantum phase without interfering its release mechanism.}
		
	\end{abstract}
	
	\maketitle
	
A bound electron can be released from atoms or molecules by absorbing photons. The photoelectron carries information on its initial bound state as well as on its emission process in the field of the parent ion. The strong impact of the residual ion on the outgoing electrons has been demonstrated over a wide range of driving photon energies, spanning from Compton scattering at high photon energies \cite{melzerRoleCoulombPotential2024}, molecular photoionization at intermediate photon energies \cite{ristMeasuringPhotoelectronEmission2021}  to strong-field ionization at very long wavelengths  \cite{brabecCoulombFocusingIntense1996,popruzhenkoStrongFieldApproximation2008,quanClassicalAspectsAboveThreshold2009,liuOriginUnexpectedLow2010,yanLowEnergyStructuresStrong2010}. In the latter case, the interaction between the parent ion and the outgoing electrons manifests in a maximum in the photoelectron's final momentum distribution close to zero or a spike-like structure in the energy spectrum, and the corresponding effect has been termed Coulomb-focusing in strong-field physics \cite{blagaStrongfieldPhotoionizationRevisited2009}. In a stationary picture, the impact of the ionic potential on the phase of the outgoing electron wave, the Coulomb phase, is well understood; dynamically, however, the phase evolution of an outgoing electron wave packet remains unprobed. The phase of the electron wave packet at asymptotic distances, is encoded in the interference fringes between direct outgoing electrons and rescattered ones influenced by the Coulomb potential  \cite{huismansTimeResolvedHolographyPhotoelectrons2011,meckelSignaturesContinuumElectron2014,heDirectVisualizationValence2018,poratAttosecondTimeresolvedPhotoelectron2018b,liPhotoelectronHolographicInterferometry2019,figueirademorissonfariaItAllPhases2020,xiePicometerResolvedPhotoemissionPosition2021}.This asymptotic phase is encoded and can be retrieved by using holography  \cite{meckelLaserInducedElectronTunneling2008,bianAttosecondTimeResolvedImaging2012,hicksteinDirectVisualizationLaserDriven2012,pullenImagingAlignedPolyatomic2015,blagaImagingUltrafastMolecular2012}, as well as RABBITT (reconstruction of attosecond harmonic beating by interference of two-photon transitions) \cite{paulObservationTrainAttosecond2001,mullerReconstructionAttosecondHarmonic2002,klunderProbingSinglePhotonIonization2011,gongAttosecondSpectroscopySizeresolved2022}  and HASE (holographic angular streaking of electrons)  \cite{eckartHolographicAngularStreaking2020,trabertAngularDependenceWigner2021}.

Very recently, not the asymptotic phase, but the time dependence of the phase of a free electron has been measured by using a pulsed standing light wave interferometer in a pump-probe setup with femtosecond time resolution. The underlying effect has been termed the ultrafast Kapitza-Dirac effect (UKDE)  \cite{linUltrafastKapitzaDiracEffect2024d}. It has been used to read out phase information long after the ionization event, leaving the ionization and the subsequent evolution of the electron wave packet undistorted. The underlying physics can be understood in analogy to the conventional Kapitza-Dirac effect, where a well collimated electron beam, which is thus narrow in momentum space, is deflected by stimulated Compton scattering of photons from a continuous standing light wave  \cite{kapitzaReflectionElectronsStanding1933b,freimundObservationKapitzaDirac2001c,freimundBraggScatteringFree2002,batelaanColloquiumIlluminatingKapitzaDirac2007c}. Analogously, the pulsed standing light wave acts on a non-collimated electron wave packet which is broad in momentum space. It creates a momentum-shifted replica of the wave packet through stimulated Compton scattering. The observable fringes in momentum space are created by the interference between the initial electron wave packet and its replicas shifted in momentum space by an integer number of  2$\boldsymbol{k}$ photon recoils from the standing light wave \cite{linUltrafastKapitzaDiracEffect2024d}, where $\boldsymbol{k}$ is the photon's wave vector (atomic units are used unless stated otherwise). As this replica is produced at an adjustable delay time after the electron's release, its phase evolution can be tracked in time. In the pioneering experiment, this pulsed interferometer has been applied to a freely propagating electron wave packet sensing the $E\times t$ term of its phase evolution.

In this Letter, we use this novel tool for tracking the ultrafast dynamical phase evolution of a Coulomb-focused electron wave packet in strong-field ionization at long time delays after its release from the parent ion. We observe so far unseen interference patterns that drastically differ from those of a free electron wave packet as reported in Ref. \cite{linUltrafastKapitzaDiracEffect2024d}. In the latter case, the diffraction patterns exhibit parallel fringes that are independent on the electron's momentum transverse to the grating and the fringes. In the current study we find that the diffraction fringes of the Coulomb-focused electrons are distorted into chromosome-shaped structures, where the spacing between the diffraction fringes shrinks with decreasing momentum transverse to the standing wave grating. We show that the underlying physics can be well understood by considering the scattering phase shift expressed as the classical momentum-space action $\int \boldsymbol{r}\cdot d\boldsymbol{p} $ in the dynamical phase evolution influenced by the ionic potential. Our observation is reproduced by a numerical simulation that takes the Coulomb potential into account. Hence, the standing light wave interferometer can serve as a precise tool to probe the phase evolution of an unknown electron wave packet.
	
	\begin{figure}
	\includegraphics[width=1.0\columnwidth]{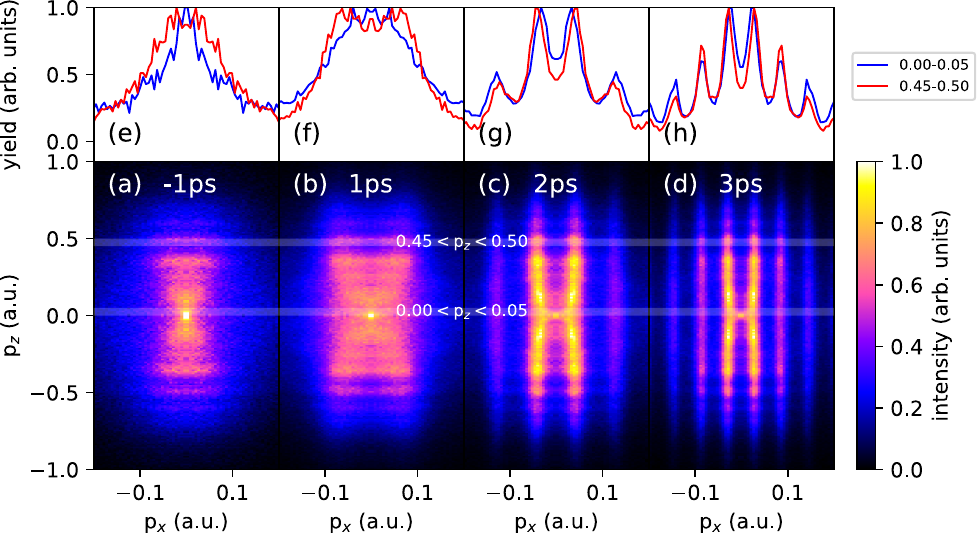}
	\caption{Time-dependent diffraction fringes for Coulomb-focused electrons. (a-d) Experimentally measured electron momentum distributions that are recorded upon strong-field ionization of xenon atoms in a pulsed standing light wave (800~nm, 60~fs, $1.0\times10^{14}$~W/cm$^2$). The ionizing pulse is followed by a weak standing light wave pulse ($0.4\times10^{14}$~W/cm$^2$) that acts as a probe pulse. Vertical axis: electron momentum along the polarization direction $p_z$; Horizontal axis: electron momentum along the light-propagation direction $p_x$, which is also the direction of the light grating. Panels (a-d) show the data taken for different time delays of -1, 1, 2, and 3 ps, respectively. (e-h) One-dimensional momentum distribution along the light-propagation axis by selecting two ranges of $p_z$ between (0.45, 0.50) and (0.00, 0.05) atomic units (shaded areas in panels (a-d)).}
	\label{Fig1}
    \end{figure}
    
Experimentally, we track the dynamical phase evolution of the electron wave packet using a pump-probe scheme. The electron wave packet is launched during the pump pulse by ionizing xenon atoms. After a variable time delay, the wave packet is probed by a standing wave, which is formed by focusing two counter-propagating laser pulses acting as a grating. Two sets of experiments are performed at different driving wavelengths of 800~nm (Coherent Legend Elite Duo, 25 fs, 800~nm, 10 kHz) and 1030~nm (Light Conversion PHAROS, 260 fs, 1030~nm, 10 kHz). The output of the PHAROS is compressed to ~50 fs by a multipass compressor (N2-Photonics). In both cases, the intensities of the probe standing wave pulses are set such that ionization is negligible. The three-dimensional momenta of the electrons were measured using a COLTRIMS reaction microscope \cite{dornerColdTargetRecoil2000a}.
	
	Figure \ref{Fig1} displays the measured electron momentum distribution along the laser polarization ($z$-axis), and propagation direction ($x$-axis) at various time delays driven by a wavelength of 800~nm. The $x$-axis is also the direction of the light grating. Using a driving wavelength of 1030~nm results in similar distributions (see Supplementary Material). The data are integrated over $p_y$. This is different from the results reported in Ref. \cite{linUltrafastKapitzaDiracEffect2024d}, where only a subset of electrons was selected for which $|p_y|>$0.1 a.u.. This gate on large $p_y$ was used to select electrons that escape the Coulomb potential so rapidly that they can be viewed as quasi-free. In the present experiment, on the contrary, we take all electrons into consideration. For reference, Figure \ref{Fig1}(a) shows the momentum distribution at negative time delay of -1 ps, when the probe pulse arrives before the pump pulse so that the electron wave packet is not scattered at the standing wave. The 2D momentum distribution is dominated by the Coulomb-focused low-energy electrons, which resemble a ``bow tie'' structure. The horizontal stripes best seen around  $|p_z|$$\sim$0.5 a.u. are a manifestation of interference of the electron wave packets released at different laser cycles, which is known as inter-cycle interference \cite{lindnerAttosecondDoubleSlitExperiment2005,arboIntracycleIntercycleInterferences2010} or above-threshold ionization \cite{agostiniFreeFreeTransitionsFollowing1979b}. Note the different scales on the horizontal and vertical axes, which masks the fact that these structures are located on circles (constant energy) in momentum space. The pattern formed by Coulomb focusing shows a strong dependence on the momentum along the polarization axis which is transverse to the direction of the standing wave. For the sake of visibility, Figure \ref{Fig1}(e) shows the 1D momentum distributions along the $p_x$-axis by selecting different slices of $p_z$ as indicated in Fig. \ref{Fig1}(a). With decreasing transverse momentum $p_z$, the bunching along the $p_x$-direction aggravates. Intuitively, trajectories of outgoing electrons launched with larger initial momentum suffer less Coulomb distortion. To obtain the dynamical phase evolution of the Coulomb-focused electrons, we apply a femtosecond standing light wave to diffract the electron wave packet at a certain time delay, as shown in Figs. \ref{Fig1}(b)-(d). Interestingly, the ``bow tie'' structure evolves into ``chromosome-like'' patterns, and the spacing between the diffraction peaks decreases as $|p_z|$ decreases. This $|p_z|$-dependence of the fringe spacing arises from the Coulomb potential of the parent ion, which modifies the dynamical phase evolution of the electron wave packet, as we will further discuss in the remainder of this paper.

	\begin{figure}
		\includegraphics[width=1.0\columnwidth]{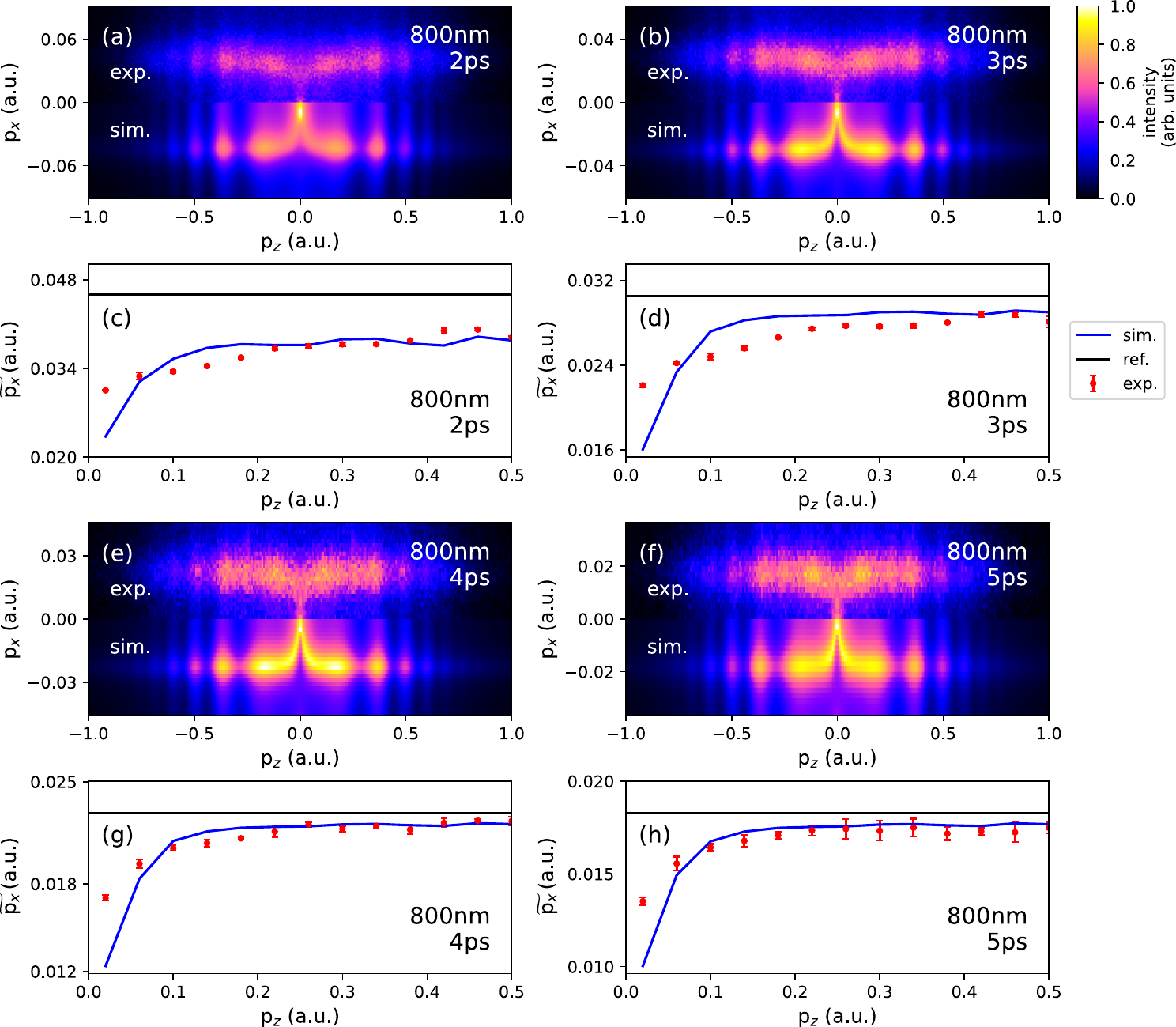}
		\caption{(a, b, e, f) Zoom into the electron momentum distributions around the first diffraction fringe at time delays of 2, 3, 4, 5 ps, respectively (standing wave made by a field with a central wavelength of 800~nm). The top half panels are experimentally measured, and the bottom half panels are simulation results. (c, d, g, h). The fringes' center $\widetilde{p_x}$ as a function of the transverse momentum $p_z$. The dots are measured values, and the curves are simulations after taking Coulomb focusing into account. The black lines indicate the fringe center for free electrons.}
		\label{Fig2}
	\end{figure}

\begin{figure}
		\includegraphics[width=1.0\columnwidth]{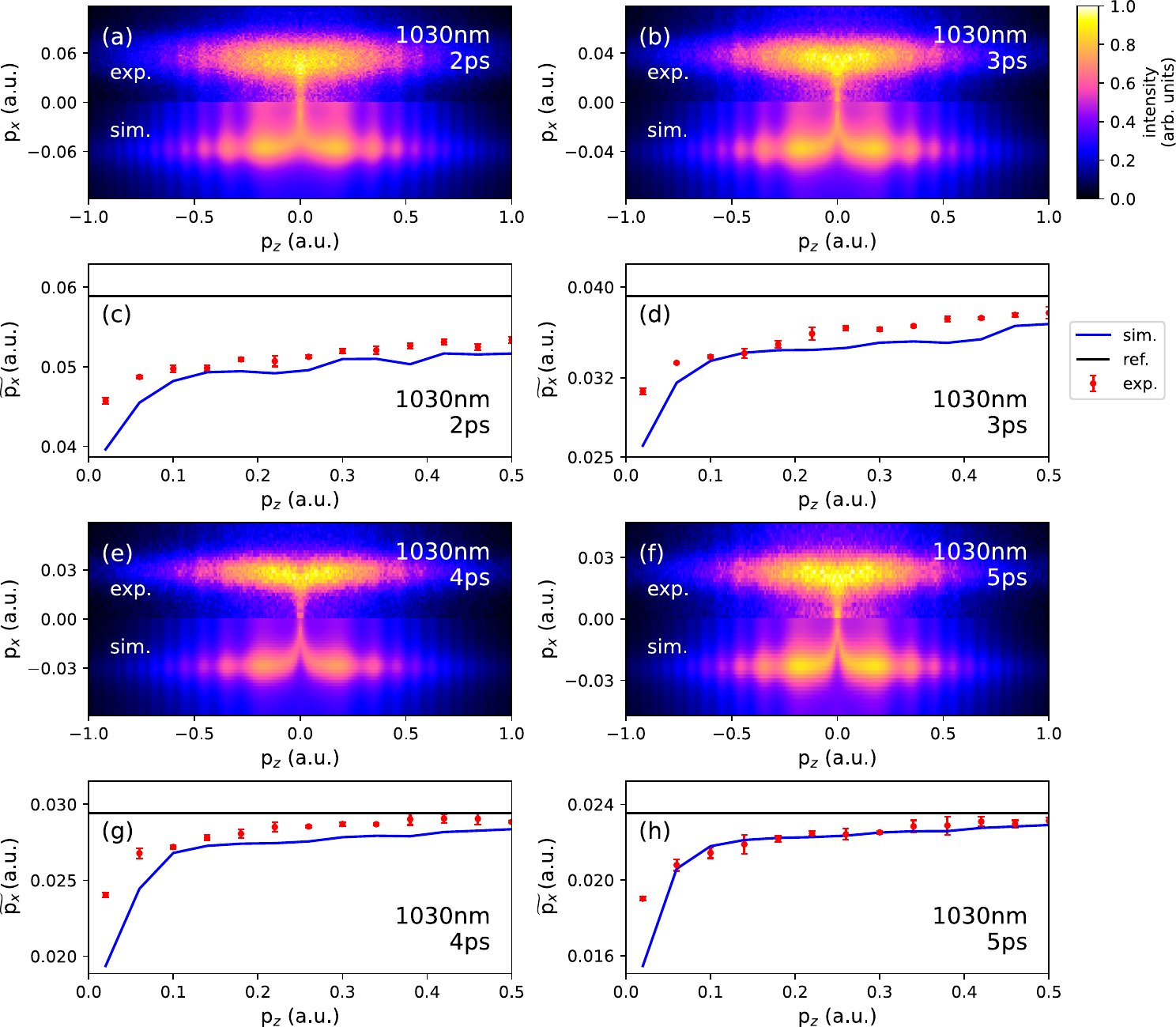}
		\caption{Similar electron momentum distributions as in Fig. \ref{Fig2}, however for a driving wavelength of 1030~nm. (a, b, e, f) time delays of 2, 3, 4, 5 ps, respectively. The top half panels are experimentally measured, and the bottom half panels are simulation results. (c, d, g, h) The fringes' center $\widetilde{p_x}$ as a function of the transverse momentum $p_z$. The dots are measured values, and the curves are simulations taking the phase shifts caused by the Coulomb potential into account. The black lines indicate the fringe center for free electrons.}
		\label{Fig3}
	\end{figure}

	Figure \ref{Fig2} shows the chromosome-shaped structures for a driving wavelength of 800 nm at time delays of 2, 3, 4 and 5 ps in more detail. In the following, we extract the position of the first fringe maxima $\widetilde{p_x}$ as a function of $p_z$, as shown in Figs. 2(c, d, g, h). For comparison, the horizontal black line shows the fringe positions for free electrons where the fringe spacing is given by $\Delta p_x=\frac{\pi}{k\tau}$, with $\tau$ being the pump-probe time delay \cite{linUltrafastKapitzaDiracEffect2024d}. Quantitatively, the deviation of the measured fringe positions from the simple case of a free electron (the black line) results from the influence of the Coulomb potential on the dynamical phase evolution of the outgoing electron wave packet. Figure \ref{Fig3} shows our experimental results at a driving wavelength of 1030~nm. The similarity between the two driving wavelengths indicates that the dynamical phase of the Coulomb-focused electrons at long time delays effectively buries the information on the ionization instant. This is in contrast to most of the previous work on holography \cite{huismansTimeResolvedHolographyPhotoelectrons2011,meckelSignaturesContinuumElectron2014,heDirectVisualizationValence2018,bianAttosecondTimeResolvedImaging2012}, which shows a strong sensitivity on the driving wavelength.
	
\begin{figure}
		\includegraphics[width=1.0\columnwidth]{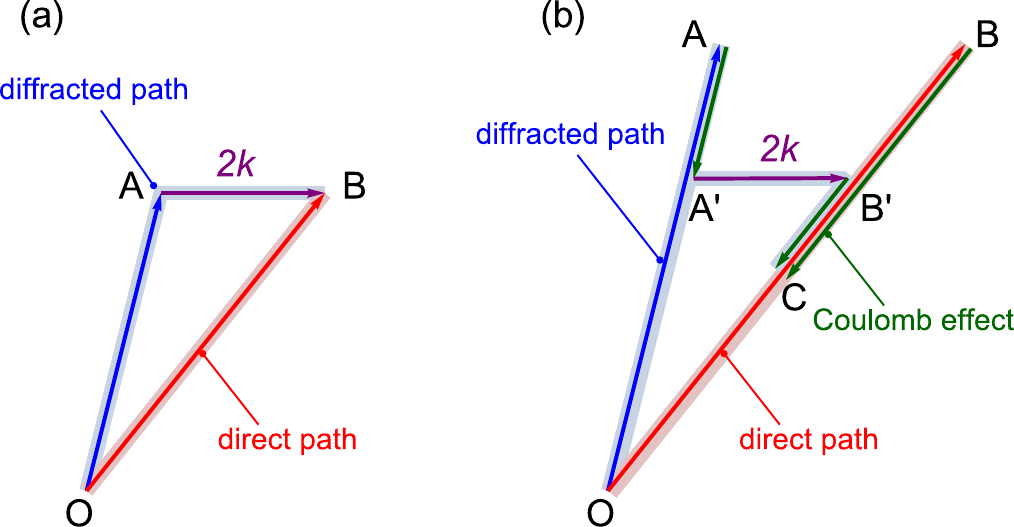}
		\caption{Interference pathways in momentum space for (a) field-free and (b) the case where Coulomb potential of the parent ion is taken into account (green vectors). Red and blue vectors represent momentum trajectories for direct and diffracted electrons, respectively. Purple vector represents the momentum absorbed from the standing wave (see text for details).}
		\label{Fig4}
	\end{figure}

	As argued above, the time-dependent fringe patterns result from the interference between the initial electron wave packet and its replicas, which have exchanged momentum of integer number of 2$\boldsymbol{k}$ with the standing wave \cite{linUltrafastKapitzaDiracEffect2024d}. For simplicity, we regard the electron creation and diffraction-mediated momentum exchange with the standing wave as instantaneous. This is well-justified as the durations of pump and probe pulses are much shorter than the time delay between them. In the Feynman path integral formalism, the phase accumulated by an electron moving along a phase space trajectory ($\boldsymbol{p}(t)$,$\boldsymbol{r}(t)$) is given by the classical action \cite{shvetsov-shilovskiSemiclassicalTwostepModel2016a}:
	 \begin{equation}
		S=\int_{t_i}^{t_f}\{-\boldsymbol{r}(t)\dot{\boldsymbol{p}}(t)-H[\boldsymbol{r}(t),\boldsymbol{p}(t)]\}dt
		\label{Equ1}
	\end{equation}
	along the corresponding classical trajectories. Solely the term that contains $\dot{\boldsymbol{p}}(t)$ in the integral is affected by the Coulomb potential. This is in contrast to the term $H[\boldsymbol{r}(t),\boldsymbol{p}(t)]$ that is only sensitive to the sum of kinetic and potential energy which is constant after the pump pulse. Figure \ref{Fig4} illustrates the two interfering pathways in momentum space. If one neglects the Coulomb potential (Fig. \ref{Fig4}(a)), the detected final momentum $\overrightarrow{OB}$ can either be part of the initial wave packet directly after the pump pulse (red arrow), or the electron with an initial momentum $\overrightarrow{OA}$ can be shifted to the detected momentum by exchange of two photon momenta after time delay $\tau$. Along both pathways the electron will move freely and its momentum will not change with position (time), thus the first integral in Eq. (\ref{Equ1}) is zero and only the energy-time term (the second integral in Eq.(\ref{Equ1})) contributes to the phase difference between the direct and the diffracted wave. Thus, the direct and the scattered electron wave packet have a phase difference of
	\begin{equation}
		\varphi_0-\varphi_+=-\frac{\boldsymbol{p}^2\tau}{2}+\frac{(\boldsymbol{p}-2\boldsymbol{k})^2\tau}{2}\approx 2\boldsymbol{k}\cdot \boldsymbol{p}\tau
		\label{Equ2}
	\end{equation}
	which is independent of the electron momentum perpendicular to photon momentum. The constructive interference condition $\varphi_0-\varphi_+=(2N+1)\pi,N\in \mathbb{Z}$ gives the peaks in momentum space (see Supplementary Material for the origin of the additional $\pi$). Consequently, Eq. (\ref{Equ2}) predicts parallel fringes spaced by $\pi/k\tau$.

However, once the Coulomb potential is involved, the first integral in Eq. (\ref{Equ1}) is nonzero and the momentum-space term plays an essential role. Figure \ref{Fig4} (b) illustrates the interfering pathways in momentum space. When no photon scattering occurs in the standing wave (direct pathway) the electron which has momentum of $\overrightarrow{OB}$ after the pump pulse decelerates to the detected momentum $\overrightarrow{OC}$ after evolving in the Coulomb potential from the parent ion. The same final momentum is reached by starting from an initial momentum of $\overrightarrow{OA}$ within the initial wave packet, decelerating to momentum  $\overrightarrow{OA'}$ in the time between the pump and the probe pulses as the electron climbs up the attractive Coulomb potential. During the probe pulse the electron changes its momentum by 2$\boldsymbol{k}$ to a momentum $\overrightarrow{OB'}$, and finally reaches the detected momentum $\overrightarrow{OC}$ by further deceleration in the Coulomb field after the probe pulse. The two pathways share the same trajectory $B'\rightarrow C$, and thus the phase difference is given by
	\begin{equation}
		\varphi_0-\varphi_+=-\int_{B}^{B'}\boldsymbol{r}\cdot d\boldsymbol{p}+\int_{A}^{A'}\boldsymbol{r}\cdot d\boldsymbol{p}-(E_0-E_+)\tau
		\label{Equ3}
	\end{equation}
	in which $E_0$ and $E_+$ represent the electron energy of two pathways before the arrival of probe standing wave. As the energy-time term alone gives parallel fringes corresponding to the Coulomb-free case, the observed chromosome-like structure unambiguously illustrates the contribution from Coulomb focusing to the phase evolution of the momentum-space term (the first term in Eq. (\ref{Equ1})).
	
	Although Eq. (\ref{Equ3}) is derived based on the simple arguments outlined above, the same expression of the phase could be reached if one computes the continuum-continuum transition matrix in detail, together with the relative amplitude of those channels (see Supplementary Material). By integrating Eq. (\ref{Equ3}) numerically, one can evaluate the final momentum distribution and compare it with experimental data. The initial wave packet $\phi_i(\boldsymbol{P})$ is fitted from Fig. \ref{Fig1}(a). Its phase is set to be momentum independent. For the purpose of the present calculation ignoring the initial phase is unproblematic as the interfering trajectories are very close in momentum, i.e. their phase difference in the initial wave packet is expected to be small. The bottom half panels in Figs. \ref{Fig2}(a, b, e, f) show the 2D momentum distributions from the numerical simulation for 800~nm, which agree well with the experimental data. The solid curves in Figs. \ref{Fig2}(c, d, g, h) are obtained by fitting the fringe center positions as a function of the transverse momentum $p_z$.The slight difference between the measurements and the simulations might come from the delta function approximation of the pump and probe standing light waves. The electron momentum distributions for the driving wavelength of 1030~nm are shown in Fig. \ref{Fig3}, which are almost the same as those for the 800~nm case. The independence of the observations on the driving wavelength further confirms the negligible difference in the initial phase upon ionization.

	In conclusion, we show that how the dynamical phase evolution of the electron wave packets under the influence of the Coulomb potential deviates from that of free ones. Phenomenologically, the diffraction patterns of the Coulomb-focused electrons turn out to be distorted to chromosome-shaped structures, while that the one of free electrons turn out to show parallel fringes. The underlying physics can be well-understood by considering the joint contribution from the momentum-space and energy-time terms in the dynamical phase evolution. Our findings demonstrate ultrafast matter wave diffraction as a precise, powerful tool to access not only the envelope of time-dependent wave packets from atoms but also their phase evolution. Future applications include phase tomography of electrons in molecular or chiral potentials.

	\begin{acknowledgments}
	K. L. acknowledges support by the Startup Funding of Zhejiang University, and the National Natural Science Foundation of China (No. 12474348).	R. D. acknowledges Funding by the European Union (ERC, Timing-FreePhase-Projectnumber 101141762). Views and opinions expressed are however those of the author(s) only and do not necessarily reflect those of the European Union or the European Research Council. Neither the European Union nor the granting authority can be held responsible for them. 
	\end{acknowledgments}

	\bibliography{Ref}
	
\end{document}